\documentstyle[aps]{revtex}

\begin{document}
\title{Radiation attenuation of bend oscillations of coherent martensitic and twin
boundaries}
\author{V.N.Dumachev\thanks{%
E-mail: dumachev@edu.vrn.ru}}
\address{Voronezh Militia Institute, Russia}
\author{V.N.Nechaev}
\address{Voronezh State Technical University, Russia}
\maketitle

\begin{abstract}
In frameworks of the macroscopic dynamic theory of coherent interfaces in
crystals the acoustic radiation of coherent boundary at passage along it of
localized wave packages is considered. It is shown that the coherent twin
boundary of acoustic waves does not radiate.
\end{abstract}

\pacs{}

The distribution lengthways interface of the localized wave packages owing
to interaction with a crystal volume is resulted of radiation friction. The
outflow of energy from coherent interphase or twin boundary is observed as
effect of acoustic radiation. The acoustic radiation caused by a wall of
mobile dislocations, is calculated and was observed by Schwencer R.O. and
Granato A.V. in [1,2]. The problem of acoustic radiation at annihilation of
dislocations was considered by Nacik V.D.and Chishko K.A. in work [3], in
which density of energy flux $P_i$ was calculated:

\[
P_i=-\sigma _{ik}\partial _tu_i, 
\]
where $u_i$ is total displacement vector; $\sigma _{ik}$ -is stress tensor.

The problem of radiation by edge dislocation of rayleigh waves is
investigated in [4]. In the present work within of the self-coordinated
macroscopic theory of domain boundaries [5, 6] the radiation field of border
is calculated. As a special case from results of our article follow the
result of work [3], about radiation of elastic waves by system of moving
dislocations.

Let's consider elastic continuum with the included defect - source of a
field of plastic deformation $s_{ik}$. The local change of a surface
position of this defect conducts to change of size of incompatibility around
of it, which in case of flat defects is shown as movement of discrepancy
dislocations. The arising at it the configuration or superficial forces
aspire to return of border of defect into habitus plane, and itself
incompatibility, owing to inertial properties of media, becomes a source of
radiation of elastic waves into volume of a material. Taking into account,
that the total deformation of a crystal $u_{ik}$, at presence of defect of
structure, is the sum of an elastic $e_{ik}$ and plastic $s_{ik}$ part: 
\[
u_{ik}=e_{ik}+s_{ik}, 
\]
then averaged on time the flow of energy radiation can be written as.

\begin{equation}
P_i=-%
{\displaystyle {1 \over T}}
\int \sigma _{ik}\partial _tu_kd{\bf r}dt=-%
{\displaystyle {1 \over T}}
\int 
{\displaystyle {d{\bf q} \over (2\pi )^3}}
{\displaystyle {d\omega  \over 2\pi }}
i\omega \lambda _{iklm}\left( iq_lu_m({\bf q},\omega )-s_{lm}\right) u_k(-%
{\bf q},-\omega ).  \label{1}
\end{equation}
where ${\bf q}$ and $\omega $ are wave vector and frequency of the elastic
vibrations; $\lambda _{iklm}=\lambda \delta _{ik}\delta _{lm}+\mu (\delta
_{il}\delta _{km}+\delta _{im}\delta _{kl})$- is elastic constants tensor in
isotropic approximation: $\lambda ,\mu $ are Lame's constants.

For a finding of a vector of total displacement we shall consider the
lagrangian of elastic continuum with any defect of structure:

\begin{equation}
L=\frac 12\int d{\bf r}dt\left( \rho \left( \partial _tu_i\right) ^2-\lambda
_{iklm}\left( \partial _iu_k-s_{ik}\right) \left( \partial
_lu_m-s_{lm}\right) \right) .  \label{2}
\end{equation}
The variation (2) with respect to $u_i$ gives the dynamic equation of the
elastic theory

\begin{equation}
\frac \delta {\delta u_i}:\rho \partial _{tt}^2u_i-\lambda _{iklm}\partial
_{kl}^2u_m=f_j({\bf r},t)  \label{3}
\end{equation}
where

\[
f_i=\lambda _{iklm}\partial _ks_{lm}. 
\]
Solving of last equation can be presented in the form

\begin{equation}
u_i({\bf r},t)=\int \int d{\bf r}^{\prime }dt^{\prime }G_{ik}({\bf r}-{\bf r}%
^{\prime },t-t^{\prime })f_k({\bf r}^{\prime },t^{\prime }),  \label{4}
\end{equation}
where

\begin{equation}
G_{ik}({\bf r},t)=\frac 1{4\pi \rho {\bf r}}\left[ \frac{n_in_k}{c_l^2}%
\delta \left( t-\frac{{\bf r}}{c_l}\right) +\frac{\delta _{ik}-n_in_k}{c_t^2}%
\delta \left( t-\frac{{\bf r}}{c_t}\right) +\frac{t\left( \delta
_{ik}-3n_in_k\right) }{{\bf r}^2}\left( \theta \left( t-\frac{{\bf r}}{c_t}%
\right) -\theta \left( t-\frac{{\bf r}}{c_l}\right) \right) \right]
\label{5}
\end{equation}
- is Green tensor of dynamic equation of elastic theory (3); $c_t,c_l$ are
sound velocity of transverse and longitudinal waves. Inserting the
Fourier-transform of the solution of equation (4):

\[
u_i({\bf q},\omega )=G_{ik}({\bf q},\omega )f_k({\bf q},\omega ) 
\]
into (1), one gets

\begin{eqnarray}
P_i &=&%
{\displaystyle {1 \over T}}
\int 
{\displaystyle {d{\bf q} \over (2\pi )^3}}
{\displaystyle {d\omega  \over 2\pi }}
\omega \lambda _{iklm}q_lf_j({\bf q},\omega )G_{jm}({\bf q},\omega )G_{nk}(-%
{\bf q},-\omega )f_n(-{\bf q},-\omega )  \label{6} \\
&&-%
{\displaystyle {1 \over T}}
\int 
{\displaystyle {d{\bf q} \over (2\pi )^3}}
{\displaystyle {d\omega  \over 2\pi }}
i\omega \lambda _{iklm}s_{lm}G_{nk}(-{\bf q},-\omega )f_n(-{\bf q},-\omega ).
\nonumber
\end{eqnarray}

It is necessary to note, that the expression (6) describes radiation of any
moving defect of continuity of structure, which capable to create a field of
plastic deformation in a material.

Let's consider coherent interface as independent object of a crystal and
according formula (6) we calculate the energy losses of wave packages which
extending lengthways it, on radiating attenuation. Writing down plastic
deformation arising at a deflection of border as

\begin{equation}
s_{ik}=\left[ S_{ik}\right] \delta (z)\zeta ({\bf r}_{\Vert },t),  \label{7}
\end{equation}
and insert it into lagrangian (2):

\[
L=\frac 12\int d{\bf r}\left( \rho \left( \partial _tu_i\right) ^2-\lambda
_{iklm}\left( \partial _iu_k-[S_{ik}]\delta (z)\zeta ({\bf r}_{\Vert
},t)\right) \left( \partial _lu_m-[S_{lm}]\delta (z)\zeta ({\bf r}_{\Vert
},t)\right) \right) , 
\]
after a variation on dynamic variables $u_i({\bf r},t)$ and $\zeta ({\bf r}%
_{\Vert },t)$ we shall receive system of the differential equations

\[
\frac \delta {\delta u_i}:\rho \partial _{tt}^2u_i-\lambda _{iklm}\partial
_{kl}^2u_m+\partial _k\left( \lambda _{ik}^s\delta (z)\zeta ({\bf r}_{\Vert
},t)\right) =0, 
\]

\[
\frac \delta {\delta \zeta }:\left( \lambda _{ik}^s\partial _iu_k-\lambda
^s\delta (z)\zeta ({\bf r}_{\Vert },t)\right) _{z=0}=[S_{jk}]\{\sigma
_{jk}\}=0, 
\]
where

\[
\lambda ^s=[S_{ik}]\lambda _{iklm}[S_{lm}]=[S_{ik}]\lambda _{ik}^s; 
\]

\[
\left[ S_{ik}\right] =1/2(n_iS_k+n_kS_i) 
\]

- is plastic deformation tensor of interface; $n_i$ is unit vector of the
normal to the habit planes; $S_k$ is plastic displacement vector; $\left[
...\right] $ jump across the boundary.

If boundary orientation is $S_i=\left( S_x,0,S_z\right) $, $n_i=(0,0,n_z)$
then the solution of Fourier-transformation of this set of equations with
respect to the variable $\zeta ({\bf r}_{\Vert },t)$ written as dispersion
equation of coherent boundary:

\begin{eqnarray}
0 &=&%
{\displaystyle {S_x^2 \over S^2}}
\left( 
{\displaystyle {q_y^2-\omega ^2/c_t^2 \over \sqrt{q_{\Vert }^2-\omega ^2/c_t^2}}}
+%
{\displaystyle {4q_y^2 \over \omega ^2/c_t^2}}
\left( \sqrt{q_{\Vert }^2-\omega ^2/c_l^2}-\sqrt{q_{\Vert }^2-\omega ^2/c_t^2%
}\right) \right)  \label{8} \\
&&  \nonumber \\
&&+%
{\displaystyle {S_z^2 \over S^2}}
\left( 
{\displaystyle {q_{\Vert }^2\left( q_{\Vert }^2-\omega ^2/c_t^2\right)  \over \omega ^2/c_t^2}}
\left( 
{\displaystyle {1 \over \sqrt{q_{\Vert }^2-\omega ^2/c_t^2}}}
-%
{\displaystyle {1 \over \sqrt{q_{\Vert }^2-\omega ^2/c_l^2}}}
\right) -%
{\displaystyle {\omega ^2/4c_t^2 \over \sqrt{q_{\Vert }^2-\omega ^2/c_l^2}}}
\right)  \nonumber
\end{eqnarray}

The roots of dispersion equation (8) determine a ratio between an own wave
vector $k_{\Vert }^2$ and own frequency of bend oscillations $\Omega $ of
coherent interface [7].

For research of radiating losses we shall write down tensor of plastic
deformation as a running wave

\[
s_{lm}=[S_{lm}]\delta (z)\zeta _{\circ }\exp (i{\bf k}_{\Vert }{\bf r}%
_{\Vert }-i\Omega t). 
\]

Then for a Fourier-image of density of force we shall receive expression

\[
f_j({\bf q},\omega )=i\zeta _{\circ }(2\pi )^3\delta ({\bf q}_{\Vert }-{\bf k%
}_{\Vert })\delta (\omega -\Omega )\lambda _{jklm}q_k[S_{lm}] 
\]

inserting which in (5) we obtain

\begin{eqnarray}
P_i &=&L^2\int 
{\displaystyle {dq_z \over 2\pi }}
\Omega \zeta _{\circ }^2\lambda _{iklm}\lambda _{nstu}\lambda
_{oprv}G_{mn}G_{ko}q_lq_sq_p[S_{tu}][S_{rv}]  \label{9} \\
&&+L^2\int 
{\displaystyle {dq_z \over 2\pi }}
\Omega \zeta _{\circ }^2\lambda _{iklm}\lambda
_{nstu}G_{nk}q_s[S_{lm}][S_{tu}],  \nonumber
\end{eqnarray}
where $L^2$ -is area of boundary.

The expression (9) allows to find all three components of a vector of a
ipulse flux, however components the parallel of habitus plane determine a
ipulse flux of an itself extending wave package. We are interested by that
part of a flux, which is radiated perpendicularly to border, into volume of
a material, that is component $P_3:$

\begin{eqnarray}
P_3 &=&L^2\int 
{\displaystyle {dq_z \over 2\pi }}
\Omega \zeta _{\circ }^2[S_{33}][S_{13}]\left( 
\begin{array}{c}
2\lambda \mu ^2\left( q_jq_1q_3G_{jk}G_{3k}+q_jq_3^2G_{jk}G_{1k}\right)
+4\mu ^3\left( q_3^3G_{3k}G_{1k}+q_3^2q_1G_{3k}G_{3k}\right) \\ 
\\ 
+\left( \lambda \mu ^2+\lambda ^2\mu \right) \left( \left(
q_1G_{3k}+q_3G_{1k}\right) q_kq_jG_{j3}+\left( q_1G_{33}+q_3G_{13}\right)
q_nq_kG_{nk}\right) \\ 
\\ 
+\left( 2\mu ^3+2\lambda \mu ^2\right) \left(
q_3^2q_kG_{13}G_{3k}+q_3^2q_kG_{33}G_{1k}+2q_1q_3q_kG_{33}G_{3k}\right) \\ 
\end{array}
\right)  \label{10} \\
&&\ \ +L^2\int 
{\displaystyle {dq_z \over 2\pi }}
\Omega \zeta _{\circ }^2[S_{33}][S_{13}]\left( \lambda \mu \left(
G_{33}q_1+G_{31}q_3+G_{n1}q_n\right) +\mu ^2\left(
2G_{33}q_1+4G_{13}q_3\right) \right) .  \nonumber
\end{eqnarray}

The further calculations we shall carry out in isotropic approximation.
Taking into account that

\[
G_{ij}=%
{\displaystyle {1 \over \mu (q^2-\omega ^2/c_t^2)}}
\left( \delta _{ij}-%
{\displaystyle {(1-\gamma ^2)q_iq_j \over q^2-\omega ^2/c_l^2}}
\right) 
\]

one rewrites (10) as

\begin{eqnarray}
P_3 &=&L^2\int 
{\displaystyle {dq_3 \over 2\pi }}
\Omega \zeta _{\circ }^2[S_{33}][S_{13}]%
{\displaystyle {1 \over \mu ^2\left( q^2-\omega ^2/c_t^2\right) ^2}}
\label{11} \\
&&\times \left( 
\begin{array}{c}
\left( \lambda \mu ^2+\lambda ^2\mu \right) \left( q_1^2+q_2^2\right)
q_1+\left( 10\mu ^3+13\lambda \mu ^2+3\lambda ^2\mu \right) q_1q_3^2-\left(
\lambda \mu ^2+\lambda ^2\mu \right) \left( q_1^2+q_2^2\right) ^2%
{\displaystyle {q_1(1-\gamma ^2) \over q^2-\omega ^2/c_l^2}}
\\ 
\\ 
-\left( 6\mu ^3+22\lambda \mu ^2+8\lambda ^2\mu \right) (1-\gamma ^2)%
{\displaystyle {\left( q_1^2+q_2^2\right) q_1q_3^2 \over q^2-\omega ^2/c_l^2}}
-\left( 30\mu ^3+29\lambda \mu ^2+7\lambda ^2\mu \right) (1-\gamma ^2)%
{\displaystyle {q_1q_3^4 \over q^2-\omega ^2/c_l^2}}
\\ 
\\ 
+\left( 8\lambda \mu ^2+4\lambda ^2\mu \right) (1-\gamma ^2)^2 
{\displaystyle {\left( q_1^2+q_2^2\right) ^2q_1q_3^2 \over (q^2-\omega ^2/c_l^2)^2}}
+\left( 16\mu ^3+24\lambda \mu ^2+8\lambda ^2\mu \right) \left(
q_1^2+q_2^2\right) (1-\gamma ^2)^2%
{\displaystyle {q_1q_3^4 \over (q^2-\omega ^2/c_l^2)^2}}
\\ 
\\ 
+\left( 16\mu ^3+16\lambda \mu ^2+4\lambda ^2\mu \right) (1-\gamma ^2)^2%
{\displaystyle {q_1q_3^6 \over (q^2-\omega ^2/c_l^2)^2}}
\end{array}
\right)  \nonumber \\
&&  \nonumber \\
&&+L^2\int 
{\displaystyle {dq_z \over 2\pi }}
\Omega \zeta _{\circ }^2[S_{33}][S_{13}]%
{\displaystyle {1 \over q^2-\omega ^2/c_t^2}}
\left( \left( 2\lambda +2\mu \right) q_1-\lambda (1-\gamma ^2)%
{\displaystyle {q_1\left( q_1^2+q_2^2\right)  \over q^2-\omega ^2/c_l^2}}
-3\left( \lambda +2\mu \right) (1-\gamma ^2)%
{\displaystyle {q_1q_3^2 \over q^2-\omega ^2/c_l^2}}
\right) ,  \nonumber
\end{eqnarray}

where $\gamma =c_t/c_l$.

After calculation integrals the expression (11) has the following form

\begin{eqnarray}
P_3 &=&L^2\Omega \zeta _{\circ }^2[S_{33}][S_{13}]q_1\lambda 
{\displaystyle {\left( 1-\gamma ^2\right)  \over 2\left( 2\gamma ^2-1\right) }}
\times  \label{12} \\
&&\ \left( 
\begin{array}{c}
{\displaystyle {1 \over 2\gamma ^2}}
\left( 
\begin{array}{c}
{\displaystyle {\left( 2\gamma ^2-1\right) {\bf q}_{\Vert }^2 \over \sqrt{{\bf q}_{\Vert }^2-\omega ^2/c_t^2}^3}}
- 
{\displaystyle {\left( 2\gamma ^2-1\right) (1-\gamma ^2){\bf q}_{\Vert }^4\left( 2\sqrt{{\bf q}_{\Vert }^2-\omega ^2/c_t^2}+\sqrt{{\bf q}_{\Vert }^2-\omega ^2/c_l^2}\right)  \over \left( \sqrt{{\bf q}_{\Vert }^2-\omega ^2/c_t^2}+\sqrt{{\bf q}_{\Vert }^2-\omega ^2/c_l^2}\right) ^2\sqrt{{\bf q}_{\Vert }^2-\omega ^2/c_t^2}^3\sqrt{{\bf q}_{\Vert }^2-\omega ^2/c_l^2}}}
\\ 
\\ 
- 
{\displaystyle {2\left( 3\gamma ^4+5\gamma ^2-4\right) {\bf q}_{\Vert }^2 \over \left( \sqrt{{\bf q}_{\Vert }^2-\omega ^2/c_t^2}+\sqrt{{\bf q}_{\Vert }^2-\omega ^2/c_l^2}\right) ^2\sqrt{{\bf q}_{\Vert }^2-\omega ^2/c_t^2}}}
+\left( \gamma ^2+7\right) 
{\displaystyle {\left( \sqrt{{\bf q}_{\Vert }^2-\omega ^2/c_t^2}+2\sqrt{{\bf q}_{\Vert }^2-\omega ^2/c_l^2}\right)  \over \left( \sqrt{{\bf q}_{\Vert }^2-\omega ^2/c_t^2}+\sqrt{{\bf q}_{\Vert }^2-\omega ^2/c_l^2}\right) ^2}}
\\ 
\\ 
- 
{\displaystyle {(1-\gamma ^2){\bf q}_{\Vert }^2 \over \left( \sqrt{{\bf q}_{\Vert }^2-\omega ^2/c_t^2}+\sqrt{{\bf q}_{\Vert }^2-\omega ^2/c_l^2}\right) ^3}}
\left( 
{\displaystyle {4\gamma ^2{\bf q}_{\Vert }^2 \over \sqrt{{\bf q}_{\Vert }^2-\omega ^2/c_t^2}\sqrt{{\bf q}_{\Vert }^2-\omega ^2/c_l^2}}}
+8(1-\gamma ^2)\right) - \\ 
\\ 
4(1-\gamma ^2) 
{\displaystyle {\left( 2{\bf q}_{\Vert }^2-\omega ^2/c_t^2-\omega ^2/c_l^2+3\sqrt{{\bf q}_{\Vert }^2-\omega ^2/c_t^2}\sqrt{{\bf q}_{\Vert }^2-\omega ^2/c_l^2}\right)  \over \left( \sqrt{{\bf q}_{\Vert }^2-\omega ^2/c_t^2}+\sqrt{{\bf q}_{\Vert }^2-\omega ^2/c_l^2}\right) ^3}}
+ 
{\displaystyle {\left( 4\gamma ^4-\gamma ^2-3\right)  \over \left( 1-\gamma ^2\right) }}
{\displaystyle {1 \over \sqrt{{\bf q}_{\Vert }^2-\omega ^2/c_t^2}}}
\end{array}
\right) \\ 
\\ 
-\left( 
\begin{array}{c}
{\displaystyle {2 \over \sqrt{{\bf q}_{\Vert }^2-\omega ^2/c_t^2}}}
+ 
{\displaystyle {\left( 2\gamma ^2-1\right) {\bf q}_{\Vert }^2 \over \left( \sqrt{{\bf q}_{\Vert }^2-\omega ^2/c_t^2}+\sqrt{{\bf q}_{\Vert }^2-\omega ^2/c_l}\right) \sqrt{{\bf q}_{\Vert }^2-\omega ^2/c_t^2}\sqrt{{\bf q}_{\Vert }^2-\omega ^2/c_l^2}}}
\\ 
- 
{\displaystyle {3 \over \left( \sqrt{{\bf q}_{\Vert }^2-\omega ^2/c_t^2}+\sqrt{{\bf q}_{\Vert }^2-\omega ^2/c_l}\right) }}
\\ 
\end{array}
\right)
\end{array}
\right)  \nonumber \\
&&  \nonumber
\end{eqnarray}

Replacement $\omega =\xi c_tq_{\Vert }$ (as it is accepted in the theory of
rayleigh waves) final expression accept a form

\begin{eqnarray}
P_3 &=&L^2\Omega \zeta _{\circ }^2S^2\lambda \sin (2\theta )\cos (\varphi )%
{\displaystyle {\left( 1-\gamma ^2\right)  \over 4\left( 2\gamma ^2-1\right) }}
\label{13} \\
&&\times \left( 
\begin{array}{c}
{\displaystyle {1 \over 2\gamma ^2}}
\left( 
\begin{array}{c}
{\displaystyle {\left( 2\gamma ^2-1\right)  \over \sqrt{1-\gamma ^2\xi ^2}^3}}
+ 
{\displaystyle {\left( 4\gamma ^4-\gamma ^2-3\right)  \over \left( 1-\gamma ^2\right) }}
{\displaystyle {1 \over \sqrt{1-\xi ^2}}}
- 
{\displaystyle {\left( 2\gamma ^2-1\right) (1-\gamma ^2)\left( 2\sqrt{1-\xi ^2}+\sqrt{1-\gamma ^2\xi ^2}\right)  \over \left( \sqrt{1-\xi ^2}+\sqrt{1-\gamma ^2\xi ^2}\right) ^2\sqrt{1-\xi ^2}^3\sqrt{1-\gamma ^2\xi ^2}}}
\\ 
\\ 
- 
{\displaystyle {2\left( 3\gamma ^4+5\gamma ^2-4\right)  \over \left( \sqrt{1-\xi ^2}+\sqrt{1-\gamma ^2\xi ^2}\right) ^2\sqrt{1-\xi ^2}}}
+\left( \gamma ^2+7\right) 
{\displaystyle {\left( \sqrt{1-\xi ^2}+2\sqrt{1-\gamma ^2\xi ^2}\right)  \over \left( \sqrt{1-\xi ^2}+\sqrt{1-\gamma ^2\xi ^2}\right) ^2}}
\\ 
\\ 
{\displaystyle {4\gamma ^2(1-\gamma ^2) \over \left( \sqrt{1-\xi ^2}+\sqrt{1-\gamma ^2\xi ^2}\right) ^3\sqrt{1-\xi ^2}\sqrt{1-\gamma ^2\xi ^2}}}
- 
{\displaystyle {8(1-\gamma ^2)^2 \over \left( \sqrt{1-\xi ^2}+\sqrt{1-\gamma ^2\xi ^2}\right) ^3}}
\\ 
\\ 
-4(1-\gamma ^2) 
{\displaystyle {\left( 2-\xi ^2-\gamma ^2\xi ^2+3\sqrt{1-\xi ^2}\sqrt{1-\gamma ^2\xi ^2}\right)  \over \left( \sqrt{1-\xi ^2}+\sqrt{1-\gamma ^2\xi ^2}\right) ^3}}
\end{array}
\right) \\ 
\\ 
-%
{\displaystyle {2 \over \sqrt{1-\xi ^2}}}
- 
{\displaystyle {\left( 2\gamma ^2-1\right)  \over \left( \sqrt{1-\xi ^2}+\sqrt{1-\gamma ^2\xi ^2}\right) \sqrt{1-\xi ^2}\sqrt{1-\gamma ^2\xi ^2}}}
+ 
{\displaystyle {3 \over \left( \sqrt{1-\xi ^2}+\sqrt{1-\gamma ^2\xi ^2}\right) }}
\end{array}
\right)  \nonumber
\end{eqnarray}

where $\theta =\arctan 
{\displaystyle {S_z \over S_x}}
$ is the dilatation angle, $\varphi =\arctan 
{\displaystyle {q_y \over q_x}}
$ direction of wave distribution located near to a boundary.

The analysis of the received expression shows, that the greatest acoustic
radiation will proceed from coherent interface with ($\theta =\pi /4$) at
distribution of the located wave packages in a direction of a horizontal
projection of a vector of displacement ($\varphi =0$). The coherent twin
boundary of acoustic waves does not radiate.

The above mentioned formulas allow also to calculate full energy of
radiation of elastic waves of moving coherent interphase or twin boundary:

\begin{equation}
I=%
\displaystyle \int 
\limits_0^\pi \rho \left( \partial _tu_i\right) ^2R_{\circ }^22\pi \sin
\varphi d\varphi .  \label{14}
\end{equation}

For this purpose it is necessary to insert (4), (5) into (14) previously
expansion of integrand expression into a series with respect to ${\bf r}%
^{\prime }{\bf n}$. In result we shall receive

\begin{equation}
I=\frac \rho {2\pi c_t^2}\left( \gamma ^6\frac 4{15}\left[ \partial
_{tt}^2D_{ik}^l\right] ^2+\frac 6{15}\left[ \partial _{tt}^2D_{ik}^t\right]
^2\right) +\frac{\left( 1/\gamma ^2-2\right) ^2}{\left( 4\pi \right) ^2c_t^2}%
\gamma ^6\left[ \partial _{tt}^2D_{oo}^l\right] ^2,  \label{15}
\end{equation}

where 
\[
D_{ik}^{t,l}=\int s_{ik}d{\bf r,} 
\]

the top index $t$ or $l$ determines the moment of time in which this size
undertakes $\left( t-%
{\displaystyle {R_{\circ } \over c_t}}
\right) $ or $\left( t-%
{\displaystyle {R_{\circ } \over c_l}}
\right) $. First two composed expressions (15) correspond to radiation of
twin boundary and coincide with result received in work [3]. Last composed
arises owing to acting of dilatations component of plastic deformation at
bend oscillations coherent interphase boundary of a martensite-type. To
compare expression (13) to results of the account of a pulse flux in work
[3], it is impossible owing to absence in it expression such as (13) in an
obvious form.

\end{document}